\documentclass{PoS}
\usepackage{amsfonts,amsmath,amssymb,amsbsy,bm,graphicx,makeidx,multicol,color}

\title{Neutrino motion and spin oscillations in magnetic field and matter currents}

\ShortTitle{Neutrino motion and spin oscillations in magnetic field and matter currents}

\author{\speaker{Artem Popov}$^a$, Pavel Pustoshny$^a$
and Alexander Studenikin$^{ab}$\\
\llap{$^a$}Department of Theoretical Physics, Moscow State University\\
119991 Moscow, Russia\\
\llap{$^b$} Joint Institute for Nuclear Research \\
Dubna 141980, Moscow Region, Russia\\
E-mail: \email{ar.popov@physics.msu.ru}, \email{kfrepp@gmail.com},
\email{studenik@srd.sinp.msu.ru}}

\abstract{A brief review of a neutrino oscillations effect in an external magnetic
field and matter currents is given. An ultra-relativistic neutrino propagation in moving
external media is investigated. We have found a new spin operator which
commutes with the corresponding Hamiltonian, exact wave functions
and energy spectrum are obtained.}

\FullConference{The European Physical Society Conference on High Energy Physics\\
		5-12 July, 2017\\
		Venice}

\begin{document}
\section{Introduction: neutrino spin oscillations in magnetic field and matter currents}
Massive neutrinos participate in electromagnetic interactions
(see \cite{Giunti:2014ixa} and \cite{Studenikin:2017pos137} for a review and the recent update). One of the most important phenomenon of nontrivial neutrino electromagnetic
interactions is the neutrino magnetic moment precession and the corresponding
spin oscillations in presence of external
electromagnetic fields. The later effect has been studied in numerous papers published during the several passed decades.

Within this scope the neutrino spin oscillations $\nu^{L}\Leftrightarrow \nu^{R}$ induced by the neutrino magnetic moment interaction with the transversal magnetic field ${\bf B}_{\perp}$ was first considered in \cite{Cisneros:1970nq}. Then spin-flavour oscillations $\nu^{L}_{e}\Leftrightarrow \nu^{R}_{\mu}$ in ${\bf B}_{\perp}$ in vacuum were discussed
in \cite{Schechter:1981hw}, the importance of the matter effect was emphasized in \cite{Okun:1986hi}.
The effect of the resonant amplification of neutrino spin oscillations in ${\bf B}_{\perp}$ in the presence of matter was proposed in \cite{Akhmedov:1988uk,Lim:1987tk}, the impact of the longitudinal magnetic field ${\bf B}_{||}$ was discussed in \cite{Akhmedov:1988hd}.
The neutrino spin oscillations in the presence of constant twisting magnetic field
were considered in \cite{Akhmedov:1988hd, Vidal:1990fr, Smirnov:1991ia, Akhmedov:1993sh,
Likhachev:1990ki,Dvornikov:2007aj,Dmitriev:2015ega}.

Recently a new approach to description of neutrino spin and spin-flavour oscillations in the presence of an arbitrary constant magnetic field have been also  developed  \cite{Dmitriev:2015ega,Studenikin:2016zdx}. Our approach, that is more precise  than the previously used,  is based on the use of the stationary states in the magnetic field for classification of neutrino spin states,
contrary to the customary approach when the neutrino helicity states
(that are in fact are not stationary in the presence of a magnetic field) are used for this purpose.

In \cite{Egorov:1999ah} neutrino spin oscillations were considered in the presence of an arbitrary constant electromagnetic fields $F_{\mu \nu}$. A neutrino spin oscillations in the presence of the field of circular and linearly polarized electromagnetic waves and superposition of an electromagnetic wave and constant magnetic field and the corresponding resonance conditions were considered in \cite{Lobanov:2001ar, Dvornikov:2001ez,Dvornikov:2004en}. Neutrino spin precession and oscillations spin evolution problem in a more general case when the neutrino
is subjected to general types of non-derivative interactions with external fields that are
given by the Lagrangian were considered in \cite{Dvornikov:2002rs}.

Recently we consider in details \cite{Fabbricatore:2016nec} neutrino mixing and oscillations in arbitrary constant magnetic field and derived an explicit expressions for the effective neutrino magnetic moment for the flavour neutrinos in terms of the corresponding magnetic moments
introduced in the neutrino mass basis (see also \cite{Studenikin:2016zdx}).

For many years, until  2004, it was believed that a neutrino helicity precession
and the corresponding spin oscillations can be induced by the neutrino magnetic
interactions with an external electromagnetic field.
A new and very interesting possibility for neutrino spin
(and spin-flavour) oscillations engendered in presence of matter background was
proposed and investigated for first time in \cite{Studenikin:2004bu}. It
was shown that neutrino spin oscillations
can be induced not only by the neutrino interaction with a  magnetic field, as it was believed
before, but also by neutrino interactions with matter in the case when there
is a transversal matter current or matter polarization. This new effect has been
explicitly highlighted in \cite{Studenikin:2004bu, Studenikin:2004tv}.

Note that the existence of the discussed effect of neutrino spin oscillations engendered by the transversal matter current and matter polarization and its importance for astrophysical applications have been confirmed in a series of recent papers \cite{Cirigliano:2014aoa, Volpe:2015rla, Kartavtsev:2015eva, Dobrynina:2016rwy}.

For the further consideration of neutrino propagation and oscillations in matter, it might be useful to know exact wave functions and energy spectrum of neutrino in arbitrary moving external media. In the next section, we solve this problem.

\section{Modified Dirac equation and integrals of motion}
In this section we are going to obtain neutrino exact wave functions and energy spectra in an arbitrary moving matter. For the sake of simplicity suppose that media consists only of neutrons. If there are macroscopic number of neutrons on the neutrino De Broglie wavelength, interaction is described by the following effective Lagrangian
\begin{equation}\label{eff_lagrangian}
\mathcal{L}_{int} = - f_\mu \left(\overline{\nu} \gamma^\mu \frac{1+\gamma^5}{2} \nu \right),f^\mu = \frac{G_F}{\sqrt{2}} j^\mu,
\end{equation}
where $j^\mu$ is matter current:
\begin{equation}\label{current}
j^\mu = \gamma(n,n {\bm v}),
\end{equation}
${\bm v}$ matter velocity, $n$ is neutron density in the laboratory frame, and $\gamma = \frac{1}{\sqrt{1 - v^2}}$ is a Lorentz-factor.

Using this Lagrangian one can derive the following modified Dirac equation for a neutrino in an arbitrary moving matter:
\begin{equation}\label{eqn_general}
\left(i\gamma^\mu \partial_\mu - \frac{1}{2}\gamma^\mu (1+\gamma^5)f_\mu - m\right)\psi (x) = 0
\end{equation}

This equation and its exact solutions for the case of non-moving matter were obtained for the first time in the work \cite{Stu_Ter}. We, in turn, consider the case of an arbitrary moving medium.

For the further consideration it's convenient to rewrite equation (\ref{eqn_general}) in Schr\"{o}dinger-like form:
\begin{equation}\label{eqn_shr}
i\frac{\partial \psi (x)}{\partial t} = \hat{H}_{matt} \psi (x),
\end{equation}
where Hamiltonian is given by
\begin{equation}\label{hamiltonian_s_v}
\hat{H}_{matt} = -\gamma^5 \left[\hat{\bm\Sigma}{\bm p} + \frac{G_F n/\sqrt{2}}{2\sqrt{1-v^2}}(1+\gamma^5) \hat{\bm\Sigma}{\bm v} \right] + \frac{G_F n/\sqrt{2}}{2\sqrt{1-v^2}}(1+\gamma^5) + \gamma^0 m
\end{equation}

Here we use Dirac representation of gamma matrices, where $\Sigma^k = \gamma^0\gamma^k \gamma^5$. Schr\"{o}dinger formulation allows us to find integrals of motion. We consider the case of a uniform matter moving with a constant velocity. Then the four-momentum operator commutes with the Hamiltonian (\ref{hamiltonian_s_v}), and hence an energy and a momentum of a neutrino are conserved. Using commutation relations for Dirac matrices, it's easy to show that:

\begin{equation}
\left[\hat{\bm \Sigma}{\bm p},\hat{H}_{matt}\right] = i \frac{G_F n/\sqrt{2}}{\sqrt{1-v^2}}(1+\gamma^5)(\hat{\bm \Sigma}\cdot[{\bm p} \times {\bm v}]),
\end{equation}
i.e. helicity is not conserved in the general case. Particular cases with helicity conservation, such as a non-moving matter or a matter that has a velocity parallel to a neutrino momentum, were previously discussed in literature.

We are interested in the case of an arbitrary matter velocity. For our purpose we introduce a new spin operator, that commutes with the Hamiltonian (\ref{hamiltonian_s_v}). That operator has the following form

\begin{equation}\label{spin_operator}
\hat{s}_v = \frac{1}{p}\left[\hat{\bm \Sigma}{\bm p} + \frac{G_F n/\sqrt{2}}{2\sqrt{1-v^2}}(1+\gamma^5) \hat{\bm \Sigma}{\bm v} \right], p = |{\bm p}|
\end{equation}
One can show that

\begin{equation}\label{spin_operator_commutator}
\left[ \hat{s}_v, \hat{H}_{matt} \right] = \frac{G_F n/\sqrt{2}}{2\sqrt{1-v^2}} \frac{m}{p} \hat{\bm\Sigma}{\bm v} [\gamma^5,\gamma^0].
\end{equation}
Therefore, in ultra-relativistic limit the new spin operator is an integral of motion.

Unlike an ordinary helicity operator, our spin operator has four eigenvalues:
\begin{equation}\label{spin_eigenstates}
s_v = \pm 1; \pm \frac{1}{p} \left|{\bm p} + \frac{G_F n/\sqrt{2}}{\sqrt{1-v^2}} {\bm v}\right|.
\end{equation}
But below we show that only two of them correspond to real physical states.

\section{Neutrino wave function and energy spectrum in moving matter}
As we previously stated, for the case of a uniform matter moving with a constant velocity, an energy and a momentum are conserved. Then for neutrino stationary states we get
\begin{equation}\label{plane_wave}
\psi (x) = \frac{e^{-i(Et - {\bm p}{\bm r})}}{L^{\frac{3}{2}}} u(p),
\end{equation}
where $u(p)$ is independent of time and spatial coordinates, and $L$ is the normalization length. Substituting this into equation (\ref{eqn_general}), we get
\begin{equation}
Eu(p) = \left[ -\gamma^5 s_v p + \frac{G_F n/\sqrt{2}}{2\sqrt{1-v^2}}(1+\gamma^5) \right]u(p)
\end{equation}
Upon the condition that the system has a nontrivial solution, we arrive to the energy spectrum:
\begin{equation}\label{spectrum}
E_{\varepsilon, s_v} = \frac{G_F n/\sqrt{2}}{2\sqrt{1-v^2}} + \varepsilon \left(s_v p - \frac{G_F n/\sqrt{2}}{2\sqrt{1-v^2}}\right),
\end{equation}
where $\varepsilon = \pm 1$.

Taking into account (\ref{spectrum}), one can transform $u(p)$  to the following form
\begin{equation}\label{u}
u(p) = \begin{pmatrix}
         \varphi \\
         \varepsilon \varphi
       \end{pmatrix},
\end{equation}
were $\varphi$ is an arbitrary two-component spinor.

We use the fact that we are looking for stationary states with a certain spin number (\ref{spin_eigenstates}):
\begin{equation}
\hat{s}_v \psi (x) = s_v \psi (x).
\end{equation}
This condition is equivalent to the following eigenvector problem:
\begin{equation}
{\bm \sigma}{\bm P}\varphi = s_v p \varphi,
\end{equation}
where
\begin{equation}
{\bm P} = {\bm p} + (1 - \varepsilon)\frac{G_F n/\sqrt{2}}{2\sqrt{1-v^2}}{\bm v}.
\end{equation}
It has nontrivial solutions only upon the condition ${\bm P}^2 = s_v^2 p^2$. Taking into account an explicit form of vector ${\bm P}$, for $s_v$ we get
\begin{equation}
s_v = \pm \frac{1}{p}\left| {\bm p} + (1-\varepsilon)\frac{G_F n/\sqrt{2}}{2\sqrt{1-v^2}}{\bm v} \right|.
\end{equation}
It means that in (\ref{spin_eigenstates}) $s_v = \pm 1$ implements only for $\varepsilon = 1$ and $s_v = \pm \frac{1}{p} \left|{\bm p} + \frac{G_F n/\sqrt{2}}{\sqrt{1-v^2}} {\bm v}\right|$ for $\varepsilon = -1$. Thus we showed, that there are only four physical states with quantum numbers $s = sgn(s_v) = \pm 1$ and $\varepsilon = \pm 1$.

Finally, we can get solutions of equation (\ref{eqn_general}):

\begin{equation}\label{solution_final}
\psi(x) = \frac{e^{-i(E_{s,\varepsilon}t - {\bm p}{\bm r})}}{2L^{\frac{3}{2}}}
\begin{pmatrix}
  \sqrt{1+s\frac{p_3+\frac{G_F n/\sqrt{2}}{2\sqrt{1-v^2}}(1-s\varepsilon)v_3}{\sqrt{{\bm p}^2 + (1-\varepsilon s)\left[\frac{G_F n/\sqrt{2}}{\sqrt{1-v^2}}{\bm p}{\bm v} + \frac{G_F^2 n^2 {\bm v}^2}{4(1-v^2)} \right]}}} \\
  s\sqrt{1-s\frac{p_3+\frac{G_F n/\sqrt{2}}{2\sqrt{1-v^2}}(1-s\varepsilon)v_3}{\sqrt{{\bm p}^2 + (1-\varepsilon s)\left[\frac{G_F n/\sqrt{2}}{\sqrt{1-v^2}}{\bm p}{\bm v} + \frac{G_F^2 n^2 {\bm v}^2}{4(1-v^2)} \right]}}} e^{i\delta_{s,\varepsilon}} \\
  s\varepsilon \sqrt{1+s\frac{p_3+\frac{G_F n/\sqrt{2}}{2\sqrt{1-v^2}}(1-s\varepsilon)v_3}{\sqrt{{\bm p}^2 + (1-\varepsilon s)\left[\frac{G_F n/\sqrt{2}}{\sqrt{1-v^2}}{\bm p}{\bm v} + \frac{G_F^2 n^2 {\bm v}^2}{4(1-v^2)} \right]}}} \\
 \varepsilon \sqrt{1-s\frac{p_3+\frac{G_F n/\sqrt{2}}{2\sqrt{1-v^2}}(1-s\varepsilon)v_3}{\sqrt{{\bm p}^2 + (1-\varepsilon s)\left[\frac{G_F n/\sqrt{2}}{\sqrt{1-v^2}}{\bm p}{\bm v} + \frac{G_F^2 n^2 {\bm v}^2}{4(1-v^2)} \right]}}}e^{i\delta_{s,\varepsilon}}
\end{pmatrix},
\end{equation}

\begin{equation}\label{spectrum_final}
E_{\varepsilon,s} = \frac{G_F n/\sqrt{2}}{2\sqrt{1-v^2}}\left(1 -s\varepsilon\right) + \varepsilon \sqrt{{\bm p}^2 + (1-\varepsilon s)\left[\frac{G_F n/\sqrt{2}}{\sqrt{1-v^2}}{\bm p}{\bm v} + \frac{G_F^2 n^2 {\bm v}^2}{4(1-v^2)} \right]},
\end{equation}

where $\tan{\delta_{s,\varepsilon}} = \frac{p_1+(1-s\varepsilon)\frac{G_F n/\sqrt{2}}{2\sqrt{1-v^2}}v_1}{p_2+(1-s\varepsilon)\frac{G_F n/\sqrt{2}}{2\sqrt{1-v^2}}v_2}$ and $s,\varepsilon = \pm 1$ are spin number and energy sign.

In the limit $n=0$ they coincide with ordinary vacuum solutions of Dirac equation. It's natural that solutions have a certain chirality, because it conserves in the limit of a massless particle. The right-chiral solutions, which correspond to $s\varepsilon = +1$, are sterile particles: right neutrino and left antineutrino.

This work was supported by the Russian Foundation for Basic Research under grants
No.~16-02-01023\,A and No.~17-52-53133\,GFEN\_a.

\end{document}